\documentclass[journal]{IEEEtran}
\usepackage{graphicx}
\usepackage{epsf}
\usepackage{mathtools}
\usepackage{amssymb}
\usepackage{amsmath, amsfonts}
\usepackage{latexsym}
\usepackage{subfigure}
\usepackage{multirow}
\usepackage{multicol}
\usepackage[table]{xcolor}
\usepackage{color}

\usepackage{tabularx}
\usepackage{lipsum}

\usepackage{algorithm}
\usepackage{algorithmic}

\usepackage{mathrsfs}

\usepackage[normalem]{ulem}

\definecolor{orange}{RGB}{255,127,0}
\definecolor{blue}{RGB}{0,0,255}
\definecolor{red}{RGB}{220,0,0}
\definecolor{green}{RGB}{0,120,0}
\definecolor{grey}{RGB}{255,120,255}

\begin{document}
{
\title{{\huge A Geometric Analysis Method for Evaluation of Coexistence between DSRC and Wi-Fi at 5.9 GHz}}

\author
{
Seungmo Kim and Carl Dietrich

\thanks{S. Kim and C. Dietrich are with the Bradley Department of Electrical and Computer Engineering, Virginia Tech in Blacksburg, VA, USA (e-mail: \{seungmo, cdietric\}@vt.edu).}
}

\maketitle

\begin{abstract}
Co-channel coexistence between Dedicated Short-Range Communications (DSRC) and Wi-Fi needs thorough study. The reason is that although the 5.850-5.925 GHz (5.9 GHz) band has been reserved for DSRC so far, the U.S. government is moving swiftly on opening the band to be shared with Wi-Fi. However, most prior work lacks sufficient scientific rigor by relying on performance metrics such as packet delivery rate (PDR) and packet delay that cannot accurately measure performance of a vehicular network that primarily uses broadcast in dissemination of packets. Precise analysis of such broadcast operation is essential for rigorous investigation of DSRC-Wi-Fi coexistence because most safety-critical applications of DSRC operate based on broadcast. This paper proposes a new metric that can more accurately characterize the performance of a broadcast-based DSRC network. The new metric is used to (i) characterize coexistence of DSRC with IEEE 802.11ac-based Wi-Fi and (ii) suggest selection of key medium access control (MAC) parameters for DSRC: inter-broadcast interval (IBI) and contention window (CW).
\end{abstract}


\IEEEpeerreviewmaketitle

\section{Introduction}\label{sec_introduction}
In August 2015, the United States Department of Transportation (DoT) released a test plan \cite{dot_testplan} for assessing the coexistence of Dedicated Short Range Communications (DSRC) and unlicensed devices in the 5.850-5.925 GHz (5.9 GHz) band. As suggested by the Congress in September 2015, the Federal Communications Commission (FCC), in its latest public notice \cite{fcc1668a1}, now seeks to refresh the record of its pending 5.9 GHz rulemaking to provide potential sharing solutions between Wi-Fi and DSRC at 5.9 GHz.

Wi-Fi provides short-range, high-speed, unlicensed wireless connections. DSRC uses short-range wireless communication links to facilitate information transfer for vehicle to infrastructure (V2I) and vehicle to vehicle (V2V), based on IEEE 802.11p for its PHY and MAC layers. The current focus of the FCC's solicitation in \cite{fcc1668a1} is two-fold: (i) prototype of interference-avoiding devices for testing; (ii) test plans to evaluate electromagnetic compatibility between unlicensed devices and DSRC.

Recent work discusses coexistence between DSRC system and Wi-Fi \cite{infocom_12}-\cite{gaurang17}. Especially the results of experiments and simulations in \cite{gaurang17} suggest a method of allocating channels for DSRC and Wi-Fi. However, despite the thorough study, the implications on the coexistence problem are unreliable due to discrepancy between the experimental and simulation results.

This leads to a conclusion that we need a method to accurately describe the performance of a DSRC network to accomplish the coexistence test plans solicited in \cite{fcc1668a1}. Another recent study \cite{priyabrata_twc16} provides an extensive analysis for a DSRC network, yet it lacks consideration of DSRC-Wi-Fi coexistence. Moreover, we found that using the current metrics makes it difficult to precisely evaluate the performance of a DSRC network. The main objective of DSRC networking is to support safety-critical applications that utilize a functionality called basic safety message (BSM). BSMs are periodically broadcast from a vehicle, which necessitates a unique metric to compute the network-wise performance. Many of the prior studies including \cite{infocom_6}-\cite{priyabrata_tits16} rely on the typical metrics that only myopically capture the broadcast nature of DSRC networks. Examples of these metrics are packet delivery rate (PDR) and packet delay/latency.

An advanced metric that addresses this problem is proposed in \cite{vanet06}, namely inter-reception time (IRT). The typical packet latency is defined as the time spent by a successful packet to travel from its source to final destination. But this metric is not suitable to capture the performance of broadcast-based safety applications since latency is measured only for successful packets. in other words, such a latency does not capture the impact of packet losses and collisions on the latency perceived by applications that are based on periodic broadcast of packets. Defined as the time elapsed between two successive successful receptions of packets broadcast by a specific transmitter (TX) vehicle, IRT can more accurately display the performance of a BSM-based safety-critical application on a DSRC network.

\begin{figure}[t]
\vspace{-0.3in}
\centering
\includegraphics[width = 0.5\linewidth]{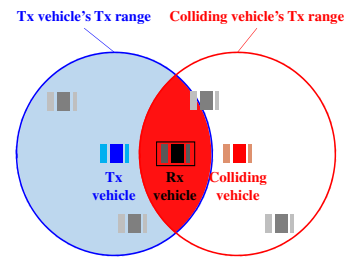}
\caption{Geometric analysis of a broadcast-based vehicular network}
\label{fig_geometry_topology}
\vspace{-0.1in}
\end{figure}

However, IRT still shows shortcomings. Limitations of IRT are highlighted when taking a geometric view to a broadcast-based vehicular network, as depicted in Fig. \ref{fig_geometry_topology}. For instance, when a packet is transmitted from a TX vehicle, all the vehicles located in the transmission range of it become potential receivers (RXs). When a packet collides with a packet from another vehicle that is at the ``left'' side of the TX vehicle, the vehicles that are located at the ``right'' side are still able to receive the packet without corruption. This geometric insight cannot be captured using the aforementioned classical metrics.

A one-dimensional geometric analysis is provided in \cite{elsevier14}. However, this paper extends the framework to a two-dimensional space, and proposes a metric that can capture partial receptions when a packet is collided. More importantly, this paper has a clear contribution to addressing the 5.9 GHz coexistence problem. We apply the proposed metric for suggesting proper values of MAC parameters when DSRC coexists with Wi-Fi--namely inter-broadcast interval (IBI) and contention window (CW).

In this paper, a new metric is therefore proposed that provides a geometric point of view to measure the performance of a broadcast-based application on a DSRC network. Based on this new method, we characterize DSRC-Wi-Fi coexistence. The contributions of this paper can be formulated as:
\begin{enumerate}
\item{This paper proposes a novel metric that evaluates the performance of a broadcast-based DSRC network. The main advantage of our method is that it can illustrate a wider perspective than the typical metrics that provide only myopic observation at a RX. Considering the broadcast nature, some RXs are still able to receive a packet from a TX even if other RXs are not. The new metric can capture such partial possibility of packet reception, which leads to a more rigorous analysis}
\item{Considering that a significant challenge that DSRC faces is co-channel coexistence with Wi-Fi, we apply the novel performance metric for characterization of DSRC-Wi-Fi coexistence. Our network simulation results suggest adequate IBI and CW values to improve the performance of a broadcast-based safety-critical application.}
\end{enumerate}

\section{System Model}\label{sec_model}
\begin{figure}[t]
\vspace{-0.3in}
\centering
\includegraphics[width = 0.7\linewidth]{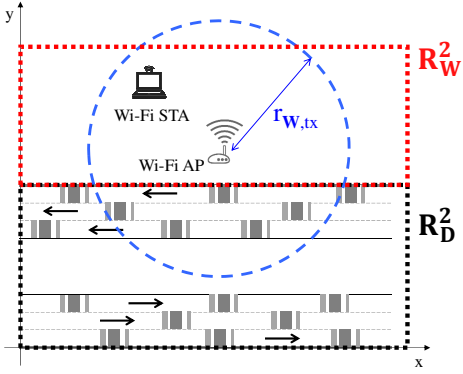}
\caption{Geometry of the simulation environment}
\label{fig_simulation_topology_5page_wifi}
\vspace{-0.1in}
\end{figure}

Fig. \ref{fig_simulation_topology_5page_wifi} illustrates the environment that comprises a 6-lane, 1000-meter (m) road with a dividing strip of 5 m between the two sides (the upper side for right-to-left and the lower side for left-to-right direction). There is a Wi-Fi system that is composed of an access point (AP) and a station (STA) right above the road, where the AP acts as the TX.

A snapshot of DSRC-Wi-Fi coexistence geometry is illustrated in Fig. \ref{fig_simulation_topology_5page_wifi} in a Cartesian coordinate system. Position of the $i$th vehicle is formally written as $\mathtt{x}_{i}=\left(x_{i},y_{i}\right) \in \mathbf{R}^2_{\boldsymbol{\mathsf{D}}}$. Since the region $\mathbf{R}_{\boldsymbol{\mathsf{D}}}^2$ is composed of 6 lanes, there are 6 values of $y_{i}$. For each lane, X axis ranges $x_{i} = \left[0, 1000\right]$, in which vehicles are distributed according to an independent and homogeneous one-dimensional Poisson Point Process (PPP) of intensity $\lambda_{dsrc}$ vehicles per 1000 m per lane. Note that for ease of analysis, we assume homogeneous traffic across all 6 lanes so that each lane has the same intensity of traffic. Thus, the number of vehicles distributed in $\mathbf{R}_{\boldsymbol{\mathsf{D}}}^2$ that is provided in our results in Section \ref{sec_evaluation} are given by $6\lambda_{dsrc}$. According to the uniformity property of a homogeneous point process \cite{daley}, we can assume that the DSRC vehicles are uniformly randomly scattered in each lane.

The DSRC network shares the 5.9 GHz spectrum with a Wi-Fi system in a co-channel manner. Position of the AP is defined as $\mathtt{x}_{ap}=\left(x_{ap},y_{ap}\right) \in \mathbf{R}^2_{\boldsymbol{\mathsf{W}}}$. Then, a Wi-Fi network is formed around the AP, which is formulated as $\left|\mathtt{x}_{ap}-\mathtt{x}_{sta}\right| \le r_{\boldsymbol{\mathsf{W}},tx}$ where $\mathtt{x}_{sta}=\left(x_{sta},y_{sta}\right)$ is position of a Wi-Fi STA, and $r_{\boldsymbol{\mathsf{W}},tx}$ is the transmission range of a Wi-Fi AP. Distribution of STAs is represented as another PPP in $\mathbf{R}_{\boldsymbol{\mathsf{W}}}^2$ with $\lambda_{sta}\left(>0\right)$.

This paper focuses on evaluating the performance of a DSRC network; the only role of Wi-Fi transmissions is external interference to the DSRC. We assume the Wi-Fi to be in a saturated situation, where the AP transmits every slot. Under this assumption, the value of $\lambda_{sta}$ does not matter as long as the AP continues transmission; thus we assume $\lambda_{sta} = 1$. Also, the Wi-Fi AP adopts an omni-directional antenna; once a position of AP is set, the only factor that determines the level of the interference at a DSRC vehicle is the vehicle's distance from the Wi-Fi AP.

There are four possible results of a packet transmission including \textit{successful delivery (DLVY)}; \textit{expiration (EXP)}; \textit{synchronized transmission (SYNC)}; and \textit{hidden node problem (HN)} \cite{elsevier14}. A DLVY does not undergo either contention nor collision. Also, with a DLVY, we assume that every RX vehicle in a TX vehicle's transmission range successfully receives the packet. On the other hand, an EXP defines a situation where a packet is not even transmitted due to contention with other vehicles or a Wi-Fi TX.

There are two main types of packet \textit{collision}: A SYNC refers to a situation where more than one TXs start transmission at the same time due to the same value of backoff in CSMA/CA. On the other hand, a HN occurs in relation to carrier-sense threshold.

\section{Proposed Performance Evaluation Method}\label{sec_rgb}
The three types of packet failure--EXP, SYNC, and HN--do not necessarily incur a lost message for all the RX vehicles in a TX vehicle's transmission range, $r_{tx}$. Based on that main idea, this section introduces a geometric method that precisely evaluates the performance of broadcast-based DSRC network.

We focus on the geometric analysis that can reflect a wider perspective of a network as opposed to gauging the performance of a certain TX-RX pair. We propose a metric that more accurately displays reception of the broadcast packets and thus represents the performance of a vehicular network more accurately. This metric, \textit{Reception Geometry for Broadcast} ($\boldsymbol{\mathsf{RGB}}$) indicator, is defined at an arbitrary TX vehicle as
\begin{align}\label{eq_rgb}
\boldsymbol{\mathsf{RGB}} &= \frac{\text{Area where a packet can be received}}{\text{Area of the transmission range}}\nonumber\\
&=\frac{\mathsf{A}_{rx}}{\pi r_{tx}^{2}}.
\end{align}
As implied in (\ref{eq_rgb}), an $\boldsymbol{\mathsf{RGB}}$ is a normalized quantity with a range of [0,1]. For discussing how to compute $\boldsymbol{\mathsf{RGB}}$, we start with DLVY and EXP since they are straightforward. As a DLVY does not undergo either contention nor collision, $\boldsymbol{\mathsf{RGB}}$ is given by
\begin{align}\label{eq_rgb_suc}
\mathsf{A}_{rx} = \pi r_{tx}^{2} \Rightarrow \boldsymbol{\mathsf{RGB}}_{dlvy}=1.
\end{align}
On the other hand, $\boldsymbol{\mathsf{RGB}}$ for an EXP is given by
\begin{align}\label{eq_rgb_exp}
\mathsf{A}_{rx} = 0 \Rightarrow \boldsymbol{\mathsf{RGB}}_{exp}=0.
\end{align}
The reason for this is that a TX vehicle cannot even start transmission once it gets backed off by a contention with other TXs in the carrier-sense range. As a result, there is no RX vehicle in the network that can receive the packet.

\begin{figure}
\vspace{-0.3in}
\subfigure[Minimum $\mathsf{A}_{rx}$]
{
\centering
\includegraphics[height = 1.5in]{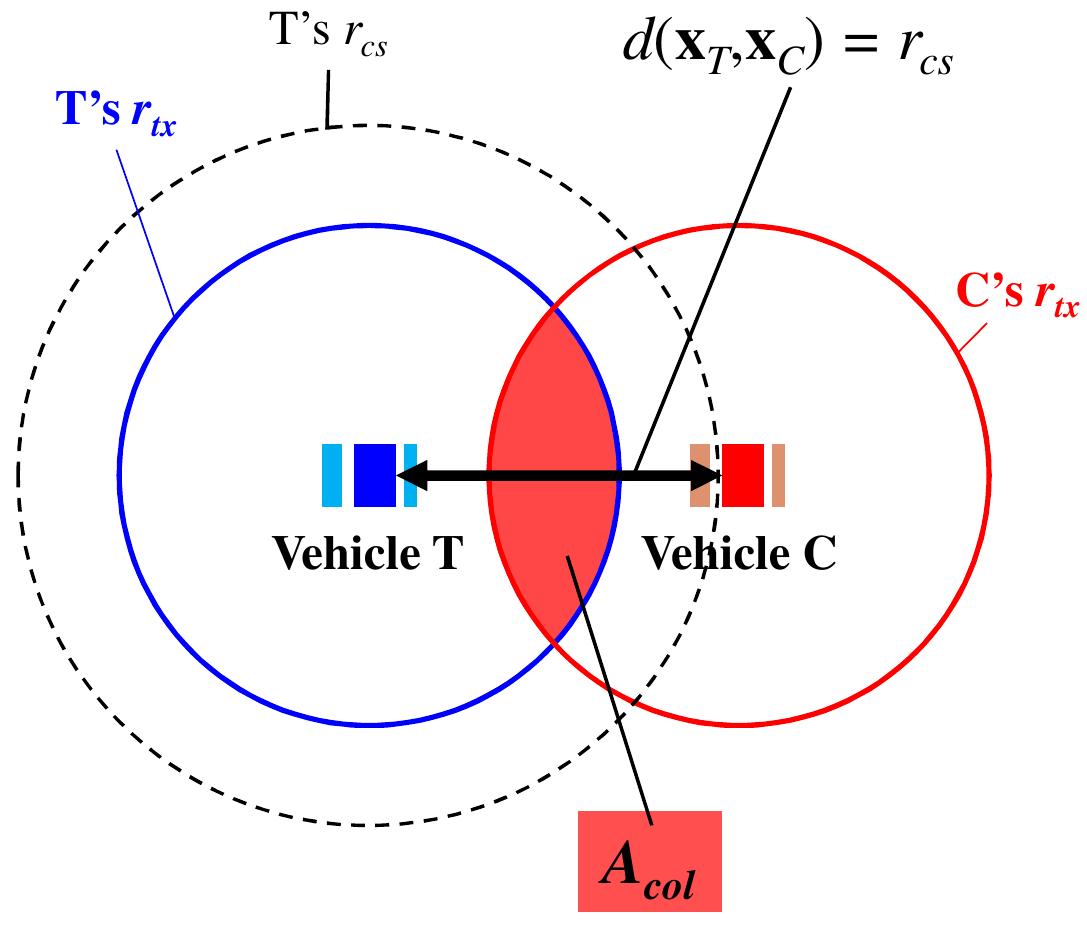}
\label{fig_geometry_sync_min}
}
\centering
\subfigure[Maximum $\mathsf{A}_{rx}$]
{
\centering
\includegraphics[height = 1.2in]{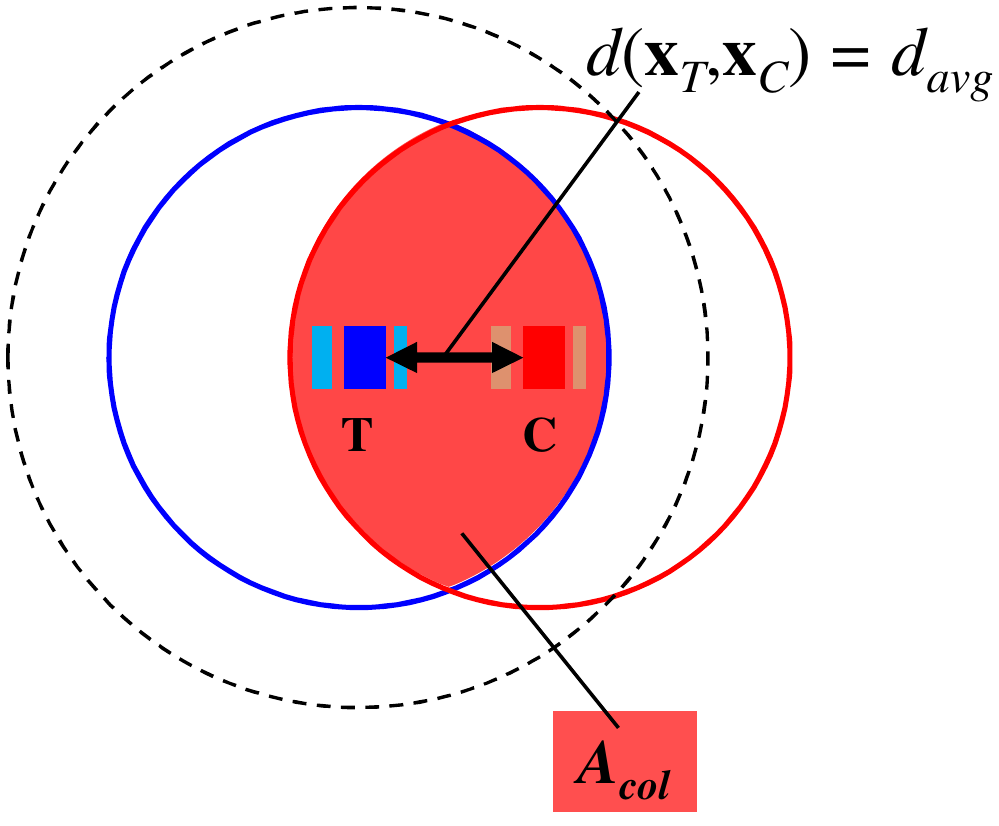}
\label{fig_geometry_sync_max}
}
\caption{Geometry of intersection for SYNC}
\label{fig_geometry_sync}
\vspace{-0.1in}
\end{figure}

Interestingly, an $\boldsymbol{\mathsf{RGB}}$ is evaluated differently between SYNC and HN. Explanation of the rationale refers to the stochastic geometry that was discussed in Section \ref{sec_model}. The geometry is illustrated in Fig. \ref{fig_geometry_sync}. A SYNC occurs when a colliding TX is in the carrier-sense range of a TX but they are assigned the same value of backoff. More specifically, $\mathsf{A}_{rx}$ of an $r_{tx}$ is determined as the ``rest'' of the intersection of two circles of transmission ranges of the TX and colliding TXs as in Fig. \ref{fig_geometry_sync}, which is given by
\begin{align}\label{eq_arxsync}
&\mathsf{A}_{rx}\left(\mathsf{d}\left(\mathtt{x}_{T},\mathtt{x}_{C}\right)\right)\nonumber\\
&= \mathsf{A}_{tx} - \mathsf{A}_{col}\left(\mathsf{d}\left(\mathtt{x}_{T},\mathtt{x}_{C}\right)\right)\nonumber\\
&= \pi r_{tx}^2 - \Biggl( 2r_{tx}^2 \cos^{-1} \left(\frac{\mathsf{d}\left(\mathtt{x}_{T},\mathtt{x}_{C}\right)}{2r_{tx}}\right)\nonumber\\
&{\rm{~~~~~~~~~~~~~~}}- \frac{\mathsf{d}\left(\mathtt{x}_{T},\mathtt{x}_{C}\right)}{2}\sqrt{4r_{tx}^2 - \mathsf{d}^2\left(\mathtt{x}_{T},\mathtt{x}_{C}\right)} \Biggr).
\end{align}
We denote by $\mathsf{A}_{col}$ the collision area and is given by $\mathsf{A}_{col} = \mathsf{A}_{tx} - \mathsf{A}_{rx}$. Also, $\mathtt{x}_{T}$ and $\mathtt{x}_{C}$ are the positions of the TX and colliding TXs, respectively. $r_{tx}$ denotes the transmission range of a vehicle. The distance between the TX vehicle and the colliding TX is defined as a function of positions $\mathtt{x}_{T}$ and $\mathtt{x}_{C}$ as
\begin{align}\label{eq_rgb_sync_inequality}
d_{avg} \le \mathsf{d}\left(\mathtt{x}_{T},\mathtt{x}_{C}\right) \le r_{cs}.
\end{align}
As Fig. \ref{fig_geometry_sync_min} shows, a $\min \mathsf{A}_{rx}$ occurs with $\max \mathsf{d}\left(\mathtt{x}_{T},\mathtt{x}_{C}\right)$; a colliding TX is placed at the border of the TX vehicle's carrier-sense range. In contrast, as in Fig. \ref{fig_geometry_sync_max}, a $\max \mathsf{A}_{rx}$ occurs with $\min \mathsf{d}\left(\mathtt{x}_{T},\mathtt{x}_{C}\right)$. Since infinitely many cases are possible for $\min d$, we consider $d_{avg}$, the average value of inter-vehicle distance, to be the minimum separation between two arbitrary vehicles. This bounds the $\boldsymbol{\mathsf{RGB}}$ for a SYNC in $\mathbf{R}_{\boldsymbol{\mathsf{D}}}^2$ as
\begin{align}\label{eq_rgb_sync}
\frac{\mathsf{A}_{rx}\left(d_{avg}\right)}{\pi r_{tx}^2} \le \boldsymbol{\mathsf{RGB}}_{sync}\left(\mathtt{x}_{T},\mathtt{x}_{C}\right) \le \frac{\mathsf{A}_{rx}\left(r_{cs}\right)}{\pi r_{tx}^2}.
\end{align}

Now, the $\boldsymbol{\mathsf{RGB}}$ for a HN can be formulated in a similar logic to (\ref{eq_arxsync})-(\ref{eq_rgb_sync}). While the way to obtain $\mathsf{A}_{rx}$ refers to (\ref{eq_arxsync}), $\mathsf{d}\left(\mathtt{x}_{T},\mathtt{x}_{C}\right)$ is computed as
\begin{align}\label{eq_rgb_hn_inequality}
r_{cs} < \mathsf{d}\left(\mathtt{x}_{T},\mathtt{x}_{C}\right) < 2r_{cs}.
\end{align}
A $\max \mathsf{d}\left(\mathtt{x}_{T},\mathtt{x}_{C}\right)$ causing $\min \mathsf{A}_{rx}$ occurs when the $r_{tx}$'s of the TX and the colliding TX contact at the border. A $\min \mathsf{d}\left(\mathtt{x}_{T},\mathtt{x}_{C}\right)$ yielding $\max \mathsf{A}_{rx}$ occurs when the colliding TX is right outside of $r_{tx}$ of the TX vehicle. This yields
\begin{align}\label{eq_rgb_hn}
\frac{\mathsf{A}_{rx}\left(r_{cs}\right)}{\pi r_{tx}^2} \le \boldsymbol{\mathsf{RGB}}_{hn}\left(\mathtt{x}_{T},\mathtt{x}_{C}\right) \le \frac{\mathsf{A}_{rx}\left(2r_{cs}\right)}{\pi r_{tx}^2}.
\end{align}

Finally, one can obtain a \textit{mean} $\boldsymbol{\mathsf{RGB}}$ that is averaged over (i) all the packets transmitted by a TX vehicle within a unit time, (ii) all the possible positions of the TX vehicle, and (iii) all the possible positions of the RX vehicles:
\begin{align}\label{eq_rgb_mean}
\overline{\boldsymbol{\mathsf{RGB}}} &= \frac{1}{\left|\mathbf{R}_{\boldsymbol{\mathsf{D}}}^2\right|^2} \int_{\mathtt{x}_{T} \in \mathbf{R}_{\boldsymbol{\mathsf{D}}}^2} \int_{\mathtt{x}_{C} \in \mathbf{R}_{\boldsymbol{\mathsf{D}}}^2} \nonumber\\
&{\rm{~~~~~~}}\sum_{k=1}^{N_{pkt}}\sum_{j \in \mathcal{S}_{pb}}{\boldsymbol{\mathsf{RGB}}_{k,j}\left(\mathtt{x}_{T},\mathtt{x}_{C}\right)}{\mathsf{P}_{k,j}} d\mathtt{x}_{T}d\mathtt{x}_{C}.
\end{align}
Denote the number of packets that are generated in a unit time by $N_{pkt}$. Alos, $\mathcal{S}_{pb}$ is the set of the four possible results of a packet transmission, $\mathcal{S}_{pb} = \left\{\text{DLVY, EXP, SYNC, HN}\right\}$. $\boldsymbol{\mathsf{RGB}}_{k,j}$ denotes each of (\ref{eq_rgb_suc}), (\ref{eq_rgb_exp}), (\ref{eq_rgb_sync}), and (\ref{eq_rgb_hn}) for the $k$th packet, and $\mathsf{P}_{k,j}$ means the probability that each of $\mathcal{S}_{pb}$ occurs.

\begin{table}[t]
\vspace{-0.3in}
\caption{Parameters}
\centering
\scriptsize
\begin{tabular}{| l | l |}
\hline
\multicolumn{1}{|c|}{\textbf{Parameter}} & \multicolumn{1}{|c|}{\textbf{Value}} \\ \hline \hline
\multicolumn{2}{|c|}{\cellcolor{gray!20}Road}\\ \hline
Length of the road & 1000 m \\ \hline
Number of lanes & 6 \\ \hline
Lane width & 4 m \\ \hline
Dividing strip width & 5 m \\ \hline
\multicolumn{2}{|c|}{\cellcolor{gray!20}DSRC}\\ \hline
Number of vehicles in $\mathbf{R}_{\boldsymbol{\mathsf{D}}}^2$ (= $6\lambda_{dsrc}$) & \{30,120,210,270\} \\ \hline
Inter-vehicle distance & \{200, 50, 28.57, 22.22\} m \\ \hline
Velocity & \{42.07, 17.49, 11.93, 9.72\} m/sec \\ \hline
Receiver sensitivity & -91 dBm\\ \hline
Transmission antenna gain & 0 dBm \\ \hline
Packet length & 500 bytes \\ \hline
Data rate & 6 Mbps \\ \hline
Slot time & 13 $\mu$s \\ \hline
\multicolumn{2}{|c|}{\cellcolor{gray!20}Wi-Fi}\\ \hline
Receiver sensitivity & -82 dBm\\ \hline
Transmission power & 30 dBm \\ \hline
CW Size & [15,1023] \\ \hline
Packet length & 10,800 bytes \\ \hline
Data rate & 54 Mbps \\ \hline
Slot time & 9 $\mu$s \\ \hline
\end{tabular}
\label{table_simulation_parameters}
\vspace{-0.1in}
\end{table}

\begin{figure*}[t]
\vspace{-0.3in}
\centering
\subfigure[With CW = 63]
{
\centering
\includegraphics[width=0.44\linewidth]{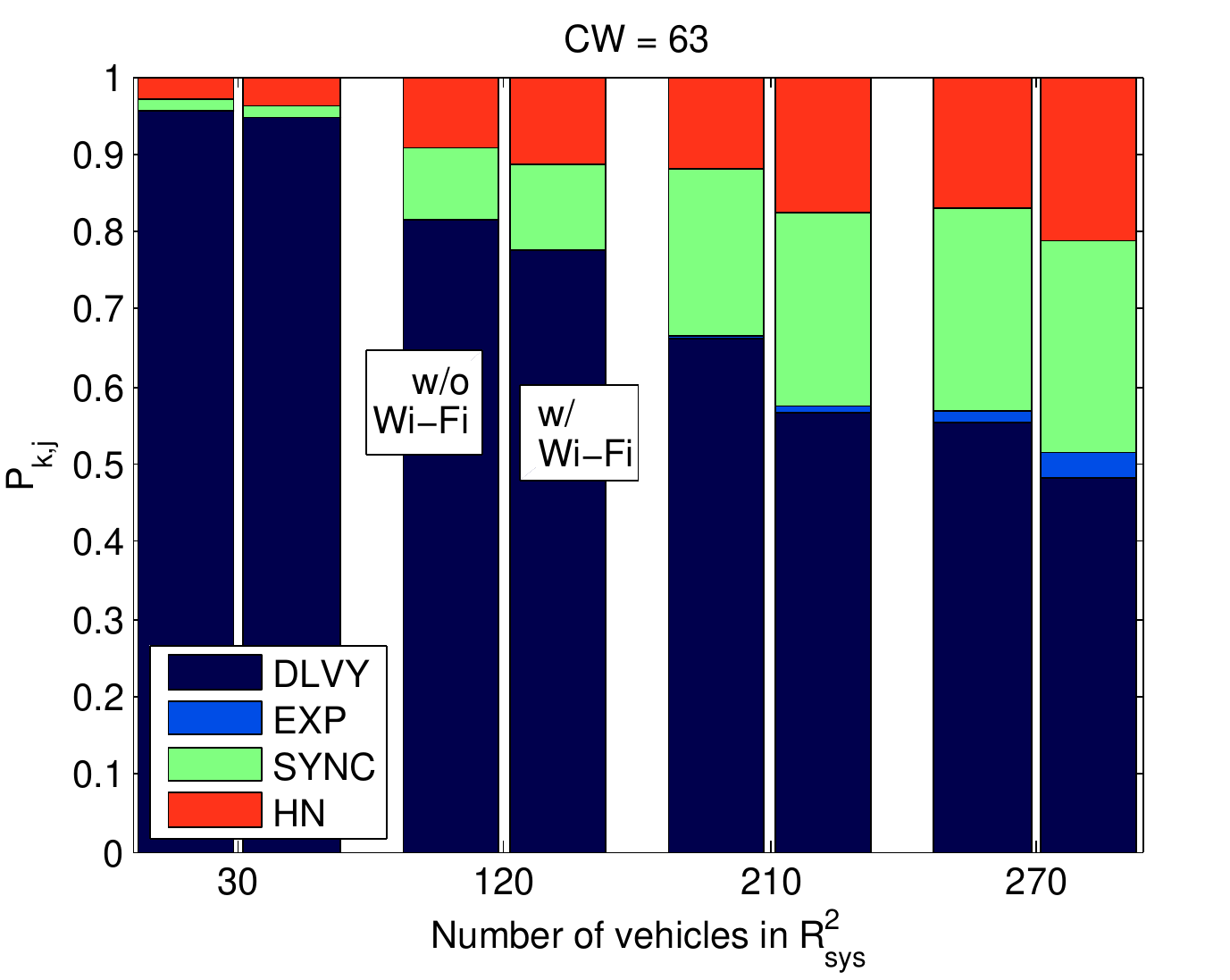}
\label{fig_proportion_63}
}
\hspace{0.1in}
\subfigure[With CW = 255]
{
\centering
\includegraphics[width=0.44\linewidth]{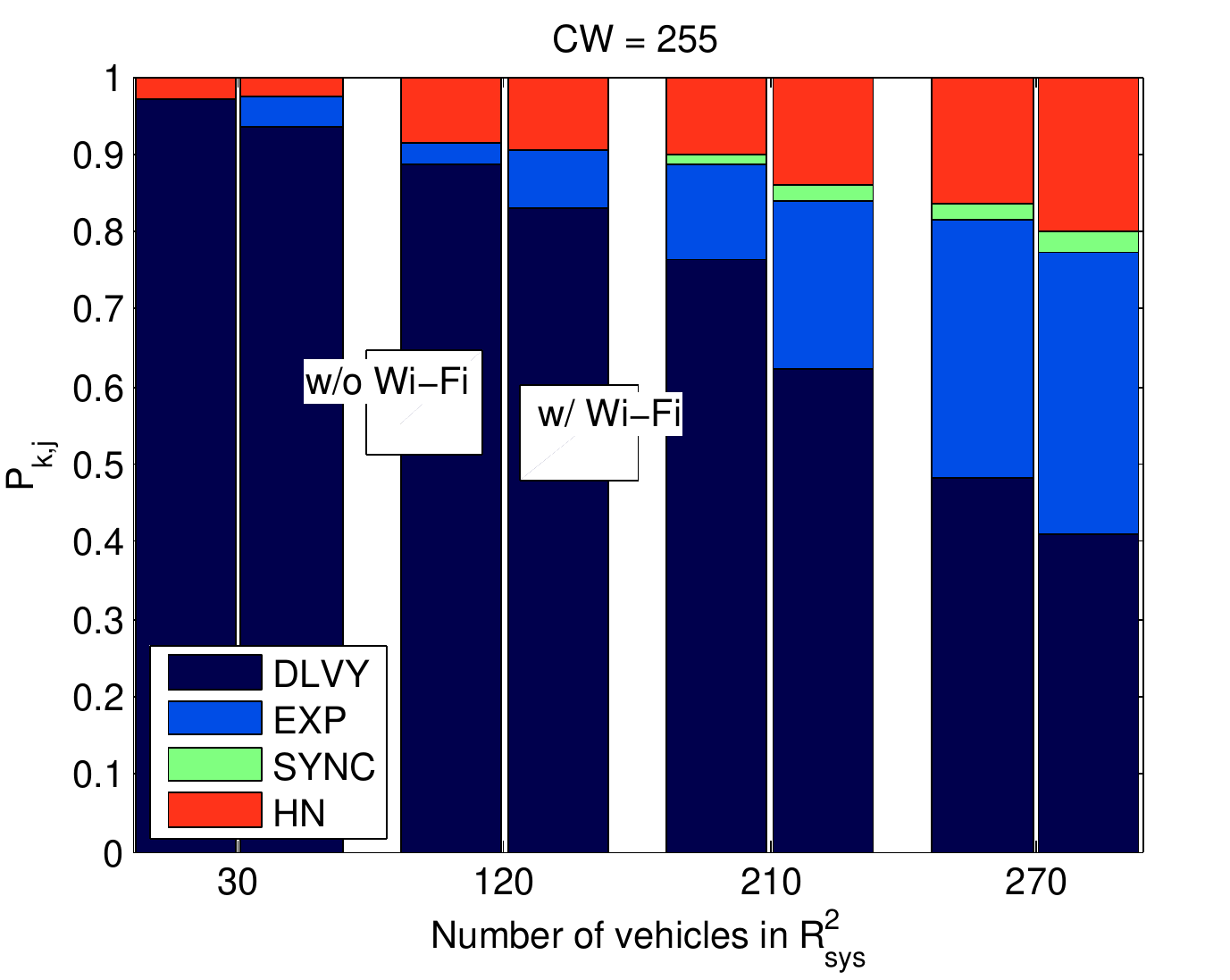}
\label{fig_proportion_255}
}
\caption{$\mathsf{P}_{j,k}$ according to vehicle density}
\label{fig_proportion}
\vspace{-0.1in}
\end{figure*}

\section{Performance Evaluation}\label{sec_evaluation}
We evaluate the performance of a DSRC-based vehicular network through the \textit{network simulator 2 (ns-2)}. The simulation topology refers to Fig. \ref{fig_simulation_topology_5page_wifi}. Each result is an average from 180 runs of 20-second experiments on \textit{ns-2}, based on the settings described in Table \ref{table_simulation_parameters}. The 180 runs are composed of 30 runs for each of the 6 lanes, in order to remove bias that could occur if not all lanes were considered.

\subsection{Setting}
As mentioned in Section \ref{sec_model}, the DSRC vehicles are uniformly distributed on the road. Their mobility setting is summarized in Table \ref{table_simulation_parameters}. The total number of vehicles situated in region $\mathbf{R}_{\boldsymbol{\mathsf{D}}}^2$ is the number of lanes multiplied by the number of vehicles per lane, i.e., $6\lambda_{dsrc}$. Inter-vehicle distance is obtained by $\lambda_{dsrc} / \text{(Road length)} = \lambda_{dsrc} / 1000$. The inter-vehicle distance accordingly determines the vehicle velocity. A set of empirical data \cite{distance_velocity} matches the velocity with the minimum inter-vehicle distance required for avoiding a collision. Since the data set was a discrete matrix that omitted some values that we needed, we fitted the data using MATLAB based on a fitting technique called \textit{shape-preserving fitter}. The resulting values of velocities were used in the \textit{ns-2} experiments.

To display the three types of packet failure--EXP, SYNC, and HN, we modified the C++ stack of the \textit{ns-2}. More specifically, we modified the packet trace function inside the stack in such a way that a colliding packet is declared to be HN/SYNC if the TX of the colliding packet is outside/within the TX range of a vehicle, respectively.

\begin{figure*}[t]
\vspace{-0.3in}
\centering
\subfigure[Without interference from Wi-Fi]
{
\centering
\includegraphics[width=0.44\linewidth]{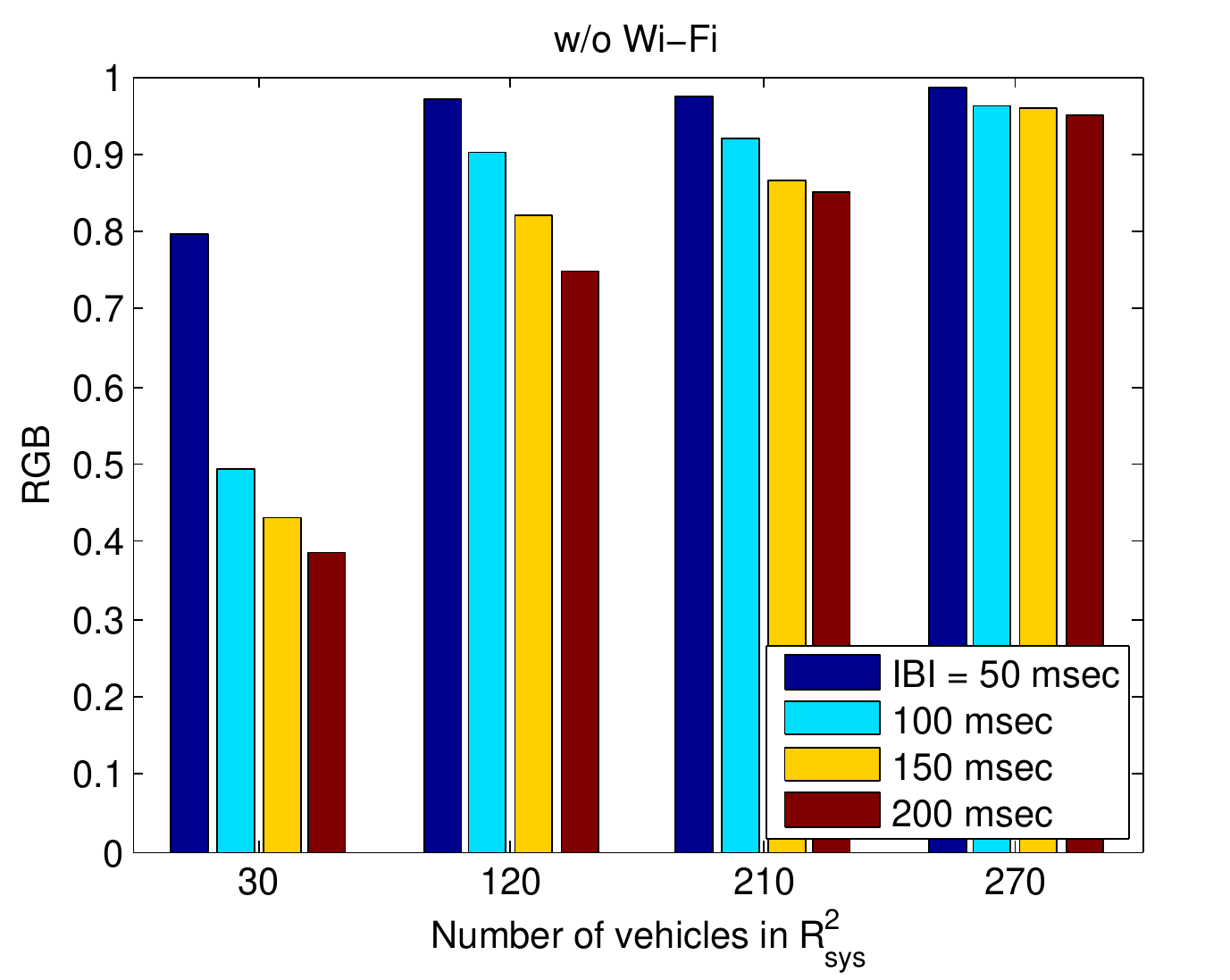}
}
\hspace{0.1in}
\subfigure[With interference from Wi-Fi]
{
\centering
\includegraphics[width=0.44\linewidth]{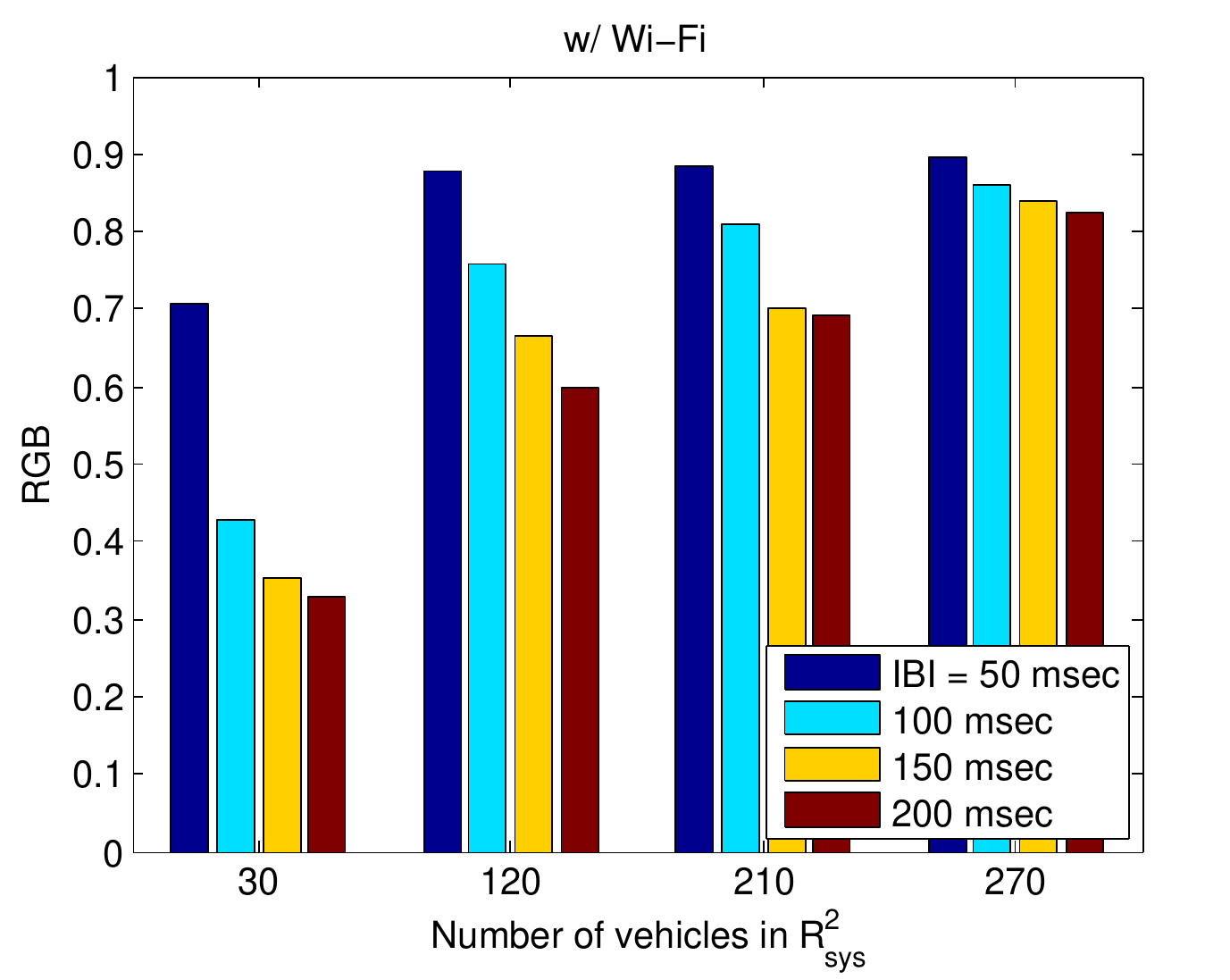}
}
\caption{$\boldsymbol{\mathsf{RGB}}$ according to IBI}
\label{fig_rgb_ibi}
\end{figure*}

\begin{figure*}[t]
\centering
\subfigure[Without interference from Wi-Fi]
{
\centering
\includegraphics[width=0.44\linewidth]{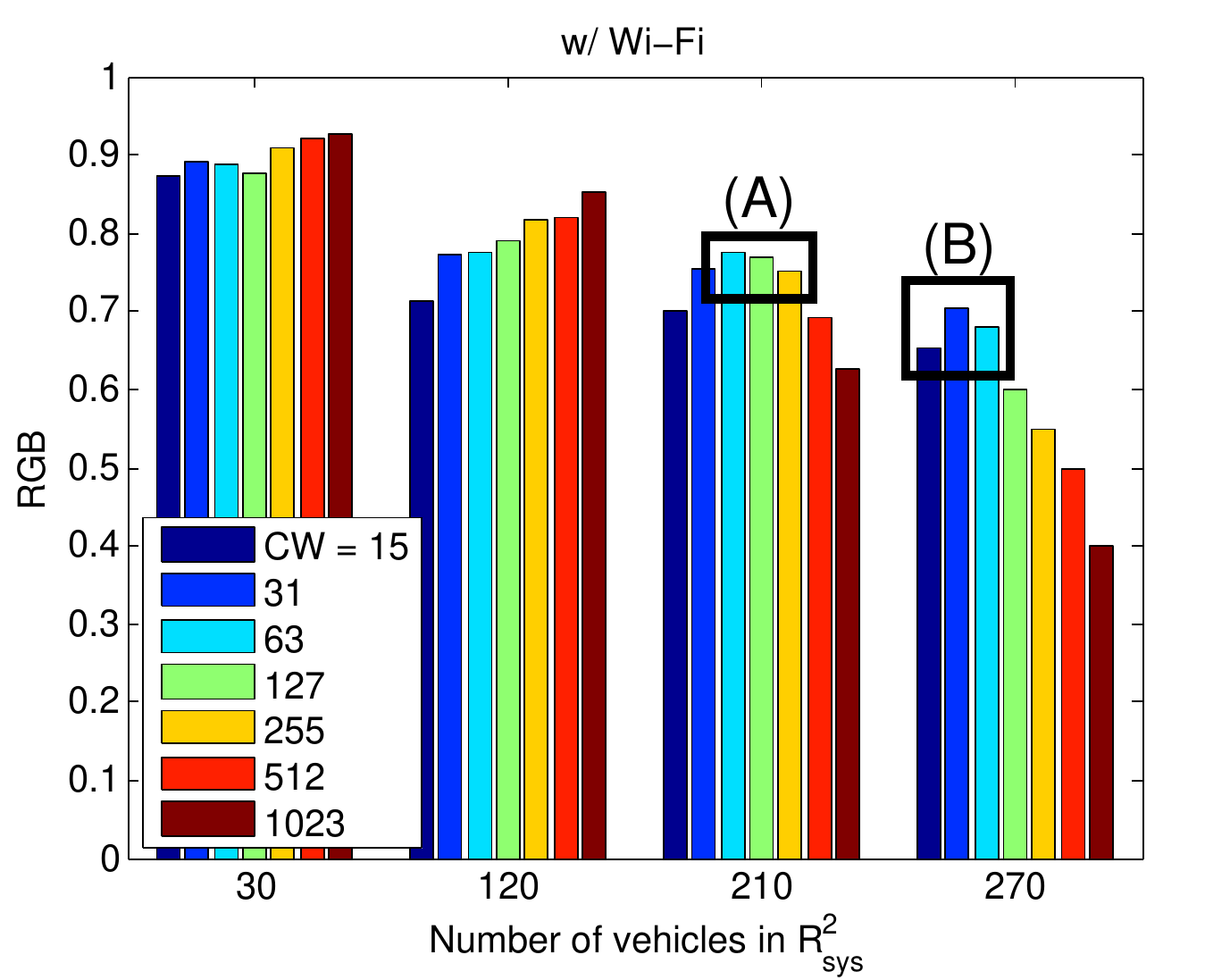}
\label{fig_rgb_cw_wo80211ac}
}
\hspace{0.1in}
\subfigure[With interference from Wi-Fi]
{
\centering
\includegraphics[width=0.44\linewidth]{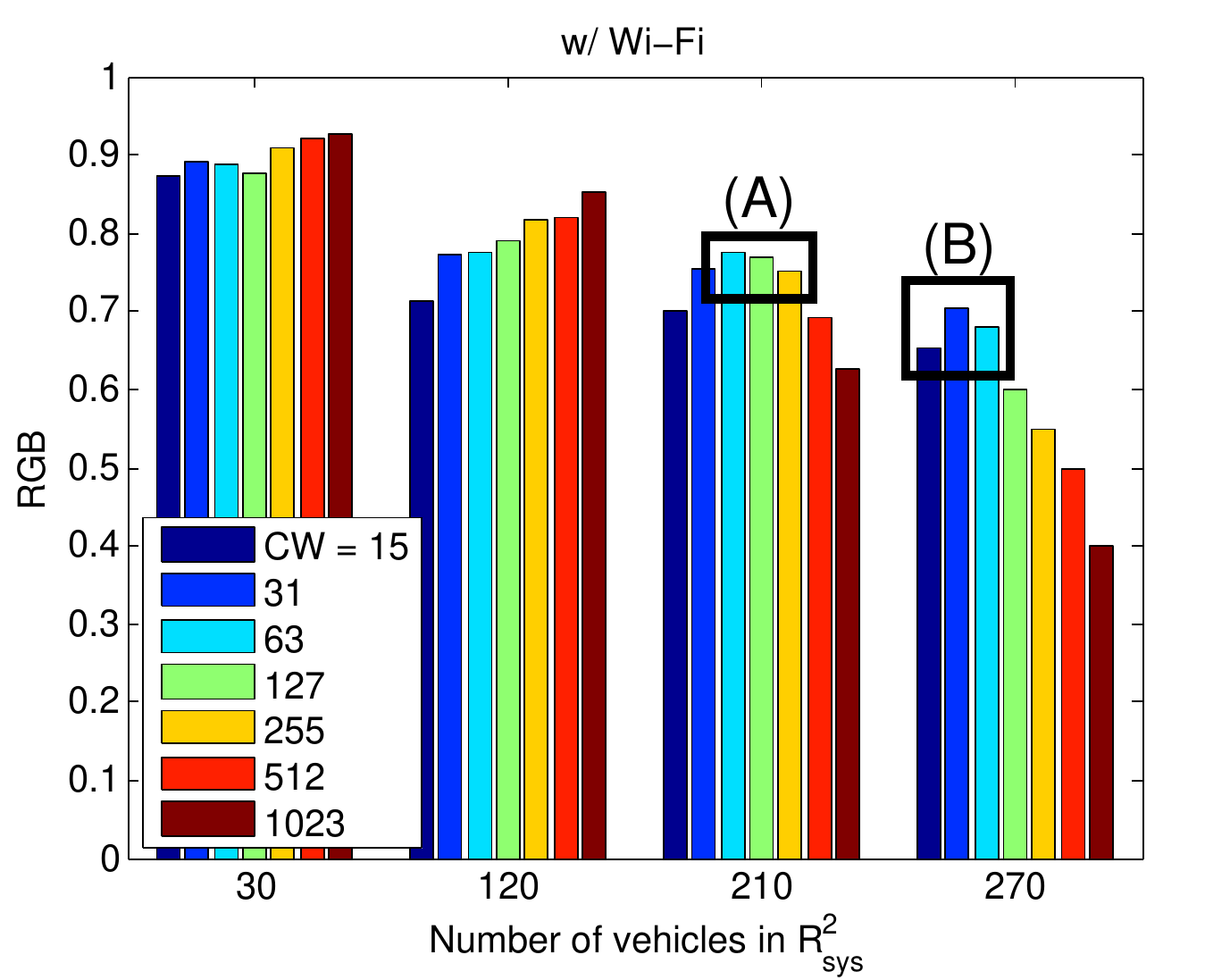}
\label{fig_rgb_cw_w80211ac}
}
\caption{$\boldsymbol{\mathsf{RGB}}$ according to CW}
\label{fig_rgb_cw}
\vspace{-0.1in}
\end{figure*}

\subsection{Results}
The results provide the values of $\mathsf{P}_{j,k}$ and $\boldsymbol{\mathsf{RGB}}_{j,k}$ that are given in (\ref{eq_rgb_mean}).

\subsubsection{$\mathsf{P}_{j,k}$ Analysis}
Fig. \ref{fig_proportion} demonstrates $\mathsf{P}_{j,k}$ for DLVY, EXP, SYNC, and HN, according to CW and presence of Wi-Fi TX. There are two bars at each value of the X-axis. The left and right bars indicate the results with and without interference from Wi-Fi. Note that comparison of Fig. \ref{fig_proportion} to Figs. 3 through 5 in \cite{elsevier14} validates our results.

Note that two representative values of CW are selected: 63 and 255. First, CW = 63 is chosen since it is a proper value to highlight impact of presence of Wi-Fi interference in a balanced manner among SYNC and HN. We found that smaller values of CW did not highlight the impact on HN due to too many SYNCs. Also, selection of CW = 255 is to compare the impact of a larger CW. In fact, Fig. \ref{fig_proportion_255} shows dramatic increase in EXP and decrease in SYNC.

Two important observations need to be made. First, EXP increases with a larger value of CW, higher traffic density, and presence of Wi-Fi TX. Second, SYNC increases with higher traffic density and presence of Wi-Fi TX, but decreases as CW increases. The reason is that now there is a wider choice of CW values for DSRC TXs and as a result $\mathsf{P}_{sync}$ gets smaller. Third, HN increases as traffic density increases and as a Wi-Fi TX is introduced. It implies that the Wi-Fi TX acts as a hidden node.

\subsubsection{$\boldsymbol{\mathsf{RGB}}$ Analysis}
Fig. \ref{fig_rgb_ibi} illustrates $\overline{\boldsymbol{\mathsf{RGB}}}$ that is given in (\ref{eq_rgb_mean}), according to (i) traffic density, (ii) IBI, and (iii) presence of interfering Wi-Fi TX. There are two noteworthy interpretations. First, the $\boldsymbol{\mathsf{RGB}}$ decreases as IBI increases. This implies that traffic congestion is a predominating factor that determines the performance of a DSRC network. When each vehicle increases IBI, the total amount of packets over the air decreases. Second, with interference from Wi-Fi, the $\boldsymbol{\mathsf{RGB}}$ is decreased for all traffic density cases.

Fig. \ref{fig_rgb_cw} shows $\overline{\boldsymbol{\mathsf{RGB}}}$ as a function of (i) traffic density, (ii) CW, and (ii) presence of Wi-Fi interference. Note the following two observations. First, a smaller value of CW is recommended as traffic density increases. This is osbserved for higher vehicle densities--210 and 270 vehicles in $\mathbf{R}_{\boldsymbol{\mathsf{D}}}^2$. Comparison of boxes (A) and (B) in Figs. \ref{fig_rgb_cw_wo80211ac} and \ref{fig_rgb_cw_w80211ac} highlights the tendency. This can be explained follows: the $\boldsymbol{\mathsf{RGB}}$ is most dependent on $\mathsf{P}_{dlvy}$, since the $\boldsymbol{\mathsf{RGB}}$ value for a DLVY is the largest (Recall that $\boldsymbol{\mathsf{RGB}}_{dlvy} = 1 > \boldsymbol{\mathsf{RGB}}_{sync}, \boldsymbol{\mathsf{RGB}}_{hn}, \boldsymbol{\mathsf{RGB}}_{exp}$). Because $\mathsf{P}_{dlvy}$ depends on both expired and collided messages, a balance needs to be found between the two quantities. After a certain point, the number of expired packets begins to prevail and the reception rate starts a slower, but steady, decrease. When $\mathsf{P}_{exp}$ goes beyond its peak (and the number of collisions starts increasing again), $\mathsf{P}_{dlvy}$ goes through another phase, where its value remains almost constant. Finally, when $\mathsf{P}_{exp}$ becomes low enough, the effect of collisions becomes predominant again and the number of total receptions starts decreasing once more. Second, Wi-Fi interference can be interpreted as additional traffic. This can be seen from the fact that the peak moves to lower CW values between Figs. \ref{fig_rgb_cw_wo80211ac} and \ref{fig_rgb_cw_w80211ac}.

\section{Conclusion}\label{sec_conclusion}
This paper proposes a new metric that accurately measures the performance of a DSRC-based vehicular network that broadcasts BSMs for safety-critical applications. Compared to traditional metrics, it more finely evaluates reception of a broadcast packet that is collided, by taking into account the RX vehicles that still are able to receive the packet. Based on the metric, it analyzes impacts of the external interference from IEEE 802.11ac-based Wi-Fi. The results suggest on the coexistence that (i) higher IBI and (ii) smaller CW are recommended in cases of higher vehicle density.



\begin{thebibliography}{99}
\setlength{\parskip}{0.001em}

\bibitem{dot_testplan} The Department of Transportation, \textit{DSRC-Unlicensed Device Test plan}, Aug. 2015.

\bibitem{fcc1668a1} Federal Communications Commission, \textit{The commission seeks to update and refresh the record in the ``unlicensed national information infrastructure (U-NII) devices in the 5 GHz band'' Proceeding}, FCC 16-68A1.

\bibitem{infocom_12} J. Lansford \textit{et al.}, ``Coexistence of unlicensed devices with DSRC systems in the 5.9 GHz band,'' in \textit{Proc. IEEE VNC 2013}.

\bibitem{infocom_13} K.-H. Chang, ``Wireless communications for vehicular safety,'' \textit{IEEE Wireless Commun.}, vol. 22, no. 1, 2015.

\bibitem{infocom_14} National Telecommunications and Information Administration (NTIA), \textit{Evaluation of the 5350-5470 MHz and 5850-5925 MHz bands}, Jan. 2013.

\bibitem{mag14} Y. Park and H. Kim, ``On the coexistence of IEEE 802.11ac and WAVE in the 5.9 GHz band,'' \textit{IEEE Commun. Mag.}, vol. 52, no. 6, 2014.

\bibitem{gaurang17} G. Naik \textit{et al.}, ``Coexistence of dedicated short range communication (DSRC) and Wi-Fi: implications to Wi-Fi performance,'' \textit{Proc. IEEE INFOCOM}, 2017.

\bibitem{priyabrata_twc16} M. Farooq \textit{et al.}, ``A stochastic geometry model for multi-hop highway vehicular communication,'' \textit{IEEE Trans. Wireless Commun.}, vol. 15, no. 3, Mar. 2016.

\bibitem{infocom_6} X. Ma and X. Chen, ``Delay and broadcast reception rates of highway safety applications in vehicular ad hoc networks,'' in \textit{Proc. IEEE Mobile Networking for Vehicular Environments}, 2007.

\bibitem{infocom_7} X. Ma \textit{et al.}, ``Performance and reliability of DSRC vehicular safety communication: a formal analysis,'' \textit{EURASIP Journal on Wireless Communications and Networking}, 2009.

\bibitem{infocom_8} X. Ma \textit{et al.}, ``Reliability analysis of one-hop safety critical broadcast services in VANETs,'' \textit{IEEE Trans. Veh. Technol.}, vol. 60, no. 8, 2011.

\bibitem{infocom_9} X. Yin \textit{et al.}, ``Performance and reliability evaluation of BSM broadcasting in DSRC with multi-channel schemes,'' \textit{IEEE Trans. Computers}, vol. 63, no. 12, 2014.

\bibitem{infocom_10} C. Campolo \textit{et al.}, ``Modeling broadcasting in IEEE 802.11p/WAVE vehicular networks,'' \textit{IEEE Commun. Lett.}, vol. 15, no. 2, 2011.

\bibitem{priyabrata_tits16} Z. Tong \textit{et al.}, ``A stochastic geometry approach to the modeling of DSRC for vehicular safety communication,'' \textit{IEEE Trans. Intell. Transp. Syst.}, vol. 17, no. 5, May 2016.

\bibitem{vanet06} T. ElBatt \textit{et al.}, ``Cooperative collision warning using dedicated short range wireless communications,'' in \textit{Proc. ACM VANET}, 2006.

\bibitem{daley} D. Daley and D. Vere-Jones, \textit{An introduction to the theory of point processes: Volume I: Elementary Theory and Methods}, Springer Probability and its Applications, Second edition, 2003.

\bibitem{elsevier14} R. Stanica \textit{et al.}, ``Reverse back-off mechanism for safety vehicular ad hoc networks,'' \textit{Elsevier Ad Hoc Networks}, vol. 16, 2014.

\bibitem{distance_velocity} http://www.arachnoid.com/lutusp/auto.html




\end{thebibliography}
\end{document}